\documentclass[aps,letterpaper,nofootinbib,preprint,showpacs,amsmath,amssymb,pdftex11pt]{revtex4-1}%

\ifnum\pdfoutput>0
	\usepackage[pdftex]{graphicx}
	\usepackage{epstopdf}
\else
	\usepackage{graphicx}%
\fi
\usepackage{latexsym}%
\usepackage[hypertex]{hyperref}%

\def\oP{{\overline P}}

\def\tR{{\widetilde R}}
\def\tF{{\widetilde F}}

\def\cF{{\mathcal F}}
\def\cH{{\mathcal H}}

\def\cR{{\mathcal R}}
\def\cW{{\mathcal W}}

\newcommand{\beq}{\begin{equation}}
\newcommand{\beqn}{\begin{equation}\nonumber}
\newcommand{\eeq}{\end{equation}}
\newcommand{\bea}{\begin{eqnarray}}
\newcommand{\bean}{\begin{eqnarray}\nonumber}
\newcommand{\eea}{\end{eqnarray}}

\begin{document}

\begin{center}
{\bf{\Large Tunneling during Quantum Collapse in AdS Spacetime}}
\bigskip\bigskip

{Cenalo Vaz$^a$\footnote{e-mail: {\tt Cenalo.Vaz@UC.Edu}} and Kinjalk Lochan$^b$\footnote{e-mail: 
{\tt kinjalk@tifr.res.in}}}

\bigskip\bigskip

{\it $^a$Department of Physics, University of Cincinnati,}\\
{\it Cincinnati, Ohio 45221-0011, USA}

\medskip

{\it $^b$Tata Institute of Fundamental Research,}\\
{\it Homi Bhabha Road, Mumbai 400 005, India}
\end{center}
\bigskip\bigskip\bigskip
\medskip

\centerline{ABSTRACT}
\bigskip\bigskip

We extend previous results on the reflection and transmission of self-gravitating dust shells across 
the apparent horizon during quantum dust collapse to non-marginally-bound dust collapse in arbitrary 
dimensions with a negative cosmological constant. We show that the Hawking temperature is independent 
of the energy function and that the wave functional describing the collapse is well behaved at the 
Hawking-Page transition point. Thermal radiation from the apparent horizon appears as a generic result 
of non-marginal collapse in AdS space-time owing to the singular structure of the Hamiltonian constraint 
at the apparent horizon.\\\\

\vfill\eject

\section{Introduction}

A fundamental expectation of a quantum theory of gravity is that it will cure the problems 
that plague classical general relativity. One hopes, for example, that singularities get resolved in quantum 
gravity \cite{cm69,bb84,kr97,ymo00,bt06}, that quantum gravity will provide the theoretcial foundation for 
cosmic censorship \cite{rp79} and that it will give a better understanding of the relationship, already 
predicted on the semi-classical level, between gravity and thermodynamics \cite{sh71,jb72,bch73,sh75}. 

In the absence of a generally agreed upon framework for such a theory, a useful approach is to quantize 
simplified classical gravitational models using canonical techniques, in the expectation that this will 
lead to new ways to look at some of the issues raised above while, at the same time, pointing to what one 
may expect out of the full theory. In this spirit, we recently developed \cite{vaz10} a novel approach to 
Hawking evaporation taking place during the collapse of a self-gravitating dust ball. This approach, 
based on an exact canonical quantization of the non-rotating, marginally bound gravity-dust 
system \cite{ltb}, exploited 
the matching conditions that must be satisfied at the apparent horizon by the wave functionals describing 
the collapse and differs from the traditional approach in which a pre-existing black hole 
is imagined to be surrounded by a tenuous field and the Bogoliubov transformation of the field operators 
is computed in the black hole background \cite{vksw02,kmsv07,fgk10}.

The geometrodynamic constraints of all LeMa\^\i tre-Tolman-Bondi (LTB) models in any dimension, with or 
without a cosmological 
constant, are expressible in terms of a canonical chart consisting of the area radius, $R$, the dust proper 
time, $\tau$, the mass function, $F$, and their conjugate momenta \cite{vws01,tsv08}. After a series of 
canonical transformations in the spirit of Kucha\v r \cite{kk94}, the Hamiltonian constraint can be shown 
to yield a Klein-Gordon-like Wheeler-DeWitt equation for the wave-functional. This equation can be solved 
by quadrature and, in the simplest cases, closed form solutions can be obtained after regularization is 
implemented on a lattice in a self-consistent manner. Self-consistency requires that the lattice 
decomposition is compatible with the diffeomorphism constraint \cite{kmv06}. In these models, the dust 
ball may be viewed as being made up of shells and the wave functional is described as the continuum limit 
of an infinite product over the shell wave functions. 

For the special case of marginal collapse with a vanishing cosmological constant in 3+1 dimensions, the 
Wheeler-DeWitt equation can be solved explicity. We showed in \cite{vaz10} that matching the shell 
wave-functions across the apparent horizon requires ingoing modes in the exterior to be accompanied by 
outgoing modes in the interior and, vice-versa, ingoing modes in the interior to be accompanied by outgoing 
modes in the exterior. In each case the relative amplitude of the outgoing wave is suppressed by the square 
root of the Boltzmann factor at a ``Hawking'' temperature given by $T_H=(4\pi F)^{-1}$, where $F$ represents
twice the mass contained within the shell. Thus the temperature varies from shell to shell, decreasing from 
the interior to the exterior, but it has the Hawking form for any given shell. 

Two separate solutions are possible: one in which there is a flow of matter toward the apparent horizon both in 
the exterior and in the interior, and another in which the flow is away from the horizon, again in both regions. 
Matter undergoing continual collapse across the apparent horizon is described by a linear superposition of these 
solutions and then, because ingoing waves in the interior are accompanied by outgoing waves in the exterior, the 
horizon appears, to the external observer with no access to the interior, to possess a reflectivity given by the 
Boltzmann factor at the above Hawking temperature. A different interpretation is also possible when the entire
shell wave functions are taken into account. Ingoing waves in the exterior must be accompanied by outgoing waves 
in the interior, whose amplitude is also suppressed by the square root of the Boltzmann factor at the Hawking 
temperature.  We showed that the transmittance of the horizon is unity, whether for waves incident from the
exterior or the interior. Thus this outgoing wave in the interior passes through the apparent horizon unhindered
but, because its amplitude is suppressed by the Boltzmann factor at the Hawking temperature relative to the
ingoing modes in the exterior, the emission probability of the horizon is given by the same factor. The net
effect is therefore reminiscent of the quasi-classical tunneling of particles through the horizon in the 
semi-classical theory \cite{kw95,sp99,gv99,pw00,mp04}.

The solutions just described relied on explicit solutions for the shell wave functions. These are available 
only in the case of the marginal models with a vanishing cosmological constant. Our aim here is to extend 
these results to non-marginally-bound LTB models in arbitrary dimension with a negative cosmological constant. 
No explicit solutions can be given in this case. Nevertheless, we will show that the results mentioned in the
previous paragraphs are indeed generic and that they are a consequence only of the essential singularity of the
Klein-Gordon equation for shells at the apparent horizon. We will then discuss how diffeomorphism invariant 
wave functionals may be reconstructed out of the shell wave functions. Hawking radiation from the apparent
horizon then appears as a consequence of the generic form of the Wheeler-DeWitt equation describing
dust collapse and not of any particular solution disucssed in the earlier work. This will provide a novel way 
to compute the entropy of the final state black hole.

The plan of this paper is as follows. In section II we recall some key results for classical dust collapse 
with a negative cosmological constant in arbitrary dimensions. In section III we present the exact wave
functional that is factorizable on a lattice and solves the Wheeler-DeWitt equation for dust collapse. 
The (collapse) wave functional can be thought of as an 
infinite product of shell wave functions, each occupying a lattice site. Matching shell wave functions across 
the horizon by analytic continuation in section IV, we argue that ingoing waves in one region must be 
accompanied by outgoing waves in the other. We superpose the two solutions to conserve the flux of shells 
across the horizon and then reconstruct the wave functional from the shell wave functions by going to the
continuum limit. A consequence of matching shells across the apparent horizon is that the amplitude for 
outgoing waves relative to ingoing ones is given by $e^{-S/2}$, where $S$ is the Bekenstein-Hawking entropy of the final state black hole. 
We close with a brief discussion of our results in section V.

\section{Non-Marginal Dust collapse with $\Lambda\neq 0$ in $d$ Dimensions}

\subsection{The classical models}

Let us begin by briefly recalling some pertinent facts about spherical dust collapse in the presence of a 
negative cosmological constant (see, for example, \cite{tgsv08} for details). The LTB models describe self-gravitating 
time-like dust whose energy momentum tensor is $T_{\mu\nu} = \varepsilon(x) U_\mu U_\nu$, where $U^\mu(\tau,\rho)$ 
is the four velocity of the dust particles which are labeled by the $\rho$ and with proper time $\tau$. The line 
element can be taken to be
\beq
ds^2 = -g_{\mu\nu} dx^\mu dx^\nu = d\tau^2 - \frac{R'^2(\tau,\rho)}{1+2 E(\rho)}d\rho^2+R^2(\tau,\rho)d\Omega_n^2
\label{ltbmetric}
\eeq
where $R(\tau,\rho)$ is the area radius, $E(\rho)$ is an arbitrary function of the shell label coordinate,
called the energy function, and $\Omega_n$ is the $n=d-2$ dimensional solid angle. Einstein's equations 
in the presence of a negative cosmological constant, which we call $-\Lambda$, 
\beq
G_{\mu\nu} + \Lambda g_{\mu\nu} = - \kappa_d T_{\mu\nu},
\eeq
yield one dynamical equation for the area radius,
\beq
{R^*}^2 = 2E(\rho) + \frac{F(\rho)}{R^{n-1}}-\frac{2\Lambda R^2}{n(n+1)},
\label{rdoteq}
\eeq
where the star refers to a derivative with respect to $\tau$ and $F(\rho)$ is a second arbitrary function of the 
shell label coordinate, called the mass function. Above, $\kappa_d$ is given in terms of the $d-$dimensional 
gravitational constant $G_d$ as $\kappa_d = 8\pi G_d$.

One also finds the energy density,
\beq
\varepsilon(\tau,\rho) = \frac n{2\kappa_d} \frac{\tF}{R^n \tR},
\label{energydensity}
\eeq
in terms of $F$, where the tilde refers to a derivative with respect to the label coordinate, $\rho$. Specific 
models are obtained by making choices of the mass and energy functions. For the solutions of \eqref{rdoteq}
to describe gravitational collapse (as opposed to an expansion) one must impose the additional condition that 
$R^*(t,r)<0$. The solutions to \eqref{rdoteq} have been explicitly given in \cite{tgsv08} and analyzed in 
detail for the marginally bound case, $E(\rho)=0$. Each shell reaches a zero area radius, $R(\tau,\rho)=0$, 
in a finite proper time, $\tau=\tau_s(\rho)$, which leads to a curvature singularity. Thus the proper time parameter lies in the interval $(-\infty,\tau_s]$. In general both naked singularities and black hole end states can form. 

Trapped surfaces occur when 
\beq
\cF~ \stackrel{\text{def}}{=}~ 1-\frac F{R^{n-1}} + \frac{2\Lambda R^2}{n(n+1)} = 0,
\eeq
which determines the physical radius, $R_h$, of the apparent horizon. $\cF$ is positive outside, {\it i.e.,} 
when $R>R_h$, and negative inside, when $R<R_h$. 

\subsection{The Canonical Formulation}

To develop a canonical formulation of the LTB models, one begins with the spherically symmetric 
Arnowitt-Deser-Misner (ADM) metric
\beq
ds^2 = N^2 dt^2 - L^2 (dr+N^r dt)^2 - R^2 d\Omega_n^2,
\eeq
where $N(t,r)$ and $N^r(t,r)$ are, respectively, the lapse and shift functions and the Einstein-Hilbert 
action for a self-gravitating dust ball
\beq
S_\text{EH} = \frac 1{2\kappa_d} \int d^d x \sqrt{-g} ({}^d\cR + 2\Lambda) -\frac 12 \int d^d x\sqrt{-g}
(g_{\alpha\beta} U^\alpha U^\beta +1),
\eeq
where $U_\alpha=-\tau_{,\alpha}$ for non-rotating dust, whete $\tau$ is the dust proper time. The phase space 
consists of the dust proper time, $\tau(t,r)$, the area radius, $R(t,r)$, the radial function, $L(t,r)$, and 
their conjugate momenta, respectively $P_\tau(t,r)$, $P_R(t,r)$ and $P_L(t,r)$. 

When the ADM metric is embedded in the spacetime described by \eqref{ltbmetric} it becomes possible, through 
a series of canonical transformations described in detail in \cite{tsv08}, to re-express the canonical constraints 
in terms of a new canonical chart consisting of the dust proper time, the area radius and the mass density function, 
$\Gamma(r)$, defined by
\beq
F(r) = \frac{2\kappa_d}{n\Omega_n}\left[M_0 + \int_0^r \Gamma(r') dr'\right]
\label{massdensity}
\eeq
and new conjugate momenta, $P_\tau(t,r)$, $\oP_R(t,r)$ and $P_\Gamma(t,r)$.  The energy function is 
expressible in this chart as
\beq
\frac 1{\sqrt{1+2E}} = \frac{2P_\tau}\Gamma
\eeq
and the transformations also absorb a boundary term, which is present in the original chart. The constraints
for the dust-gravity system in any dimension are
\bea
\cH^g &=& P_\tau^2 + \cF \oP_R^2 -\frac{\Gamma^2}{\cF} \approx 0\cr\cr
\cH_r &=& \tau'P_\tau + R' \oP_R -\Gamma P_\Gamma' \approx 0,
\label{constraints}
\eea
where the prime denotes a derivative with respect to the ADM label coordinate, $r$. The Hamiltonian constraint 
in \eqref{constraints} will be seen to contain no derivative terms, which makes it easier to quantize. 
However, the Poisson brackets of the Hamiltonian with itself vanishes, indicating that the 
Hamiltonian constraint does not generate hypersurface deformations. Rather, the transformations generated 
by the Hamiltonian constraint act along the dust flow lines.

Of importance in what follows will be the following relationship between the dust proper time and the 
remaining canonical variables (see, for example, \cite{vgks07})
\beq
\tau' = \frac{2P_\Gamma'}a \pm R' \frac{\sqrt{1-a^2 \cF}}{a\cF},
\eeq
where $a = 1/\sqrt{1+2E}$.\footnote{Einstein's equations guarantee that $1-a^2\cF >0$ for $0<a<1$, 
which corresponds to the case $E>0$.} The positive sign describes a collapsing dust cloud in the exterior 
and an expanding dust cloud in the interior whereas the negative sign describes an expanding cloud in the exterior and a collapsing cloud in the interior. Integrating on a hypersurface of constant $t$ we have 
the formal solution 
\beq
\tau = 2\int_{t=\text{const.}} \frac{dP_\Gamma}a \pm \int_{t=\text{const.}}^R dR \frac{\sqrt{1-a^2 \cF}}
{a\cF} + \widetilde{\tau}(t),
\label{pt}
\eeq
where $\widetilde{\tau}(t)$ is undetermined. This integral can be difficult to solve for arbitrary ($r$
dependent) mass and energy functions, but when they are both constant beyond some boundary, $r_b$, then 
the solution may be expressed as
\beq
a\tau = 2P_\Gamma \pm \int^R dR \frac{\sqrt{1-a^2\cF}}\cF + \widetilde{\tau}(t).
\label{ptforsads}
\eeq
In this case we are dealing with the static Schwarzschild-AdS geometry for which $2P_\Gamma$ may be associated 
with the Killing time \cite{kk94}, so \eqref{ptforsads} gives the relationship between Schwarzschild-AdS and 
Painlev\'e -Gullstrand time \cite{mp01}. Solutions in the absence of a cosmological constant have been given in
\cite{kmv06,vgks07}.

\section{Quantum States in A Lattice Decomposition}

When Dirac's quantization condition is used to raise the classical constraints to operator constraints, which 
act on a wave functional, the Hamiltonian constraint turns into the Wheeler-DeWitt equation and the momentum 
constraint imposes spatial diffeomorphism invariance on the wave functional. One sees that the second is solved 
automatically by a wave-functional of the form
\beq
\Psi[\tau,R,F] = U\left[\int_{-\infty}^\infty dr \Gamma(r) \cW(\tau(r),R(r),F(r))\right]
\label{wavefunctionaldef}
\eeq
provided that $\cW$ contains no explicit dependence on $r$ and where $U:\mathbb{R} \rightarrow \mathbb{C}$ is 
an arbitrary differentiable function of its argument. 

The wave functional is factorizable on a lattice placed on the real line (eg., see \cite{vaz10}) if $U$ is 
chosen to be the exponential map for then, taking the lattice spacing to be $\sigma$, 
\eqref{wavefunctionaldef} can be written as
\beq
\Psi[\tau,R,F] = \lim_{\sigma\rightarrow 0} \prod_i \psi_i(\tau_i,R_i,F_i)
\eeq
where 
\beq
\psi_i(\tau_i,R_i,F_i) = e^{\sigma\Gamma_i \cW(\tau_i,R_i,F_i)}
\eeq
and we have used $X_i=X(r_i)$. Thus we can think of $\Psi[\tau,R,F]$ as an infinite product of shell wave 
functions, each occupying a lattice site. Each shell wave function satisfies
\beq
\left[\hbar^2\left(\frac{\partial^2}{\partial\tau_i^2} + \cF_i \frac{\partial^2}{\partial R_i^2} + A_i 
\frac{\partial}{\partial R_i} + B_i\right) + \frac{\sigma^2\Gamma_i^2}{\cF_i}\right]\psi_i = 0
\label{kgeqn}
\eeq
where $A_i=A(R_i,F_i)$ and $B_i=B(R_i,F_i)$ are functions capturing the factor ordering ambiguities 
that are always present in the canonical approach. They can be uniquely determined by  requiring the 
above equation to be independent of the lattice spacing; one finds the general positive energy 
solutions \cite{tsv08} 
\beq
\psi_i = e^{\omega_i b_i}\times\exp\left\{-\frac{i\omega_i}\hbar \left[a_i\tau_i \pm \int^{R_i} dR_i 
\frac{\sqrt{1-a_i^2\cF_i}}{\cF_i}\right]\right\},
\label{shellwavefunctions}
\eeq
where $\omega_i = \sigma \Gamma_i/2$ and the factor $e^{\omega_i b_i}$ amounts to a normalization. Note that 
shell ``$i$'' crosses the apparent horizon when 
$\cF_i=0$, which is an essential singularity of the wave equation. From the shell wave functions in 
\eqref{shellwavefunctions} one reconstructs the wave functionals with the ansatz \eqref{wavefunctionaldef} 
and $U=\exp$ as
\beq
\Psi[\tau,R,\Gamma] = e^{\frac 12\int dr \Gamma b(F(r))}\times \exp\left\{-\frac i{2\hbar}
\int dr\Gamma\left[a(F(r))\tau\pm \int^R_{r=\text{const.}} dR \frac{\sqrt{1- a^2(F(r))\cF}}
\cF\right]\right\}\\
\label{collapsewavefunctional}
\eeq
where we have set $a(r)=a(F(r))$ and $b(r)=b(F(r))$ as is required for diffeomorphism invariance.

\section{Tunneling}

The wave-functions of the previous section are defined in the interior as well as the exterior of the 
apparent horizon, but the Wheeler-DeWitt equation has an essential singularity at the apparent horizon along 
the path of integration. In order to match interior to exterior solutions, it is necessary to deform the path 
in the complex $R$-plane. The direction of the deformation is chosen so that positive energy solutions decay. 
While this deformed path does not correspond to the trajectory of any classical particle it represents a 
tunneling of $s$-waves across the gravitational barrier represented by the apparent horizon. This is analogous 
to the quasi-classical tunneling approach employed in semi-classical analyses \cite{kw95,sp99,gv99,pw00,mp04}.

\subsection{Shell Wave Functions}

From the expression for the phase in \eqref{shellwavefunctions}, 
\beq
\cW^{(\pm)}_i = \frac{\omega_i}\hbar \left[a_i\tau_i \pm \int^{R_i} dR_i \frac{\sqrt{1-a_i^2\cF_i}}{\cF_i}
\right],
\label{phase}
\eeq
the phase velocity of the $i^\text{th}$ shell wave function is given by 
\beq
\dot R_i = \mp \frac{a_i \cF_i}{\sqrt{1-a_i^2\cF_i}}.
\eeq
Thus the positive sign in \eqref{phase} describes ingoing waves in the exterior ($\cF_i>0$), whereas it describes 
outgoing waves in the interior ($\cF_i<0$) and, likewise, the negative sign describes outgoing waves in the 
exterior and ingoing waves in the interior. 

A closed form solution for the integrals appearing in \eqref{collapsewavefunctional} and \eqref{pt} cannot be 
given when the mass function and/or the energy function and/or the cosmological constant are non-vanishing. We may 
however analyze their properties near the apparent horizon in the following way. Noting that $\cF_i = 0$, 
equivalently $R_i=R_{i,h}$, is a singularity of the integral appearing in the phase, $\cW_i$, we define the integral 
by analytically continuing to the complex plane and deforming the integration path so as to go around the pole at 
$R_{i,h}$ in a semi-circle of radius $\epsilon$ drawn in the upper half plane. Let $L_\epsilon$ denote the 
deformed path and let $S_\epsilon$ denote the semi-circle of radius $\epsilon$ around $R_{i,h}$, then
\beq
\int^{R_i} dR_i \frac{\sqrt{1-a_i^2\cF_i}}{\cF_i}~ \stackrel{\text{def}}{=}~ \lim_{\epsilon 
\rightarrow 0} {\int\limits_{L_\epsilon}}^{R_i} dR_i \frac{\sqrt{1-a_i^2\cF_i}}{\cF_i}.
\eeq
Performing the integration from left to right,\footnote{The same result is obtained if the integration 
is performed from right to left.} for $R_i = R_{i,h}+\epsilon$ we have
\beq
{\int\limits_{L_\epsilon}}^{R_{i,h}+\epsilon}dR_i \frac{\sqrt{1-a_i^2\cF_i}}{\cF_i} = 
{\int\limits_{L_\epsilon}}^{R_{i,h}-\epsilon}dR_i 
\frac{\sqrt{1-a_i^2\cF_i}}{\cF_i} + \int\limits_{S_\epsilon} dR_i \frac{\sqrt{1-a_i^2\cF_i}}{\cF_i} 
\eeq
and, for the integral over the semi-circle, a Laurent series expansion about $\cF_i=0$ gives to lowest 
order 
\beq
\int_{S_\epsilon} dR_i \frac{\sqrt{1-a_i^2 \cF_i}}{\cF_i} = \frac 1{\cF'_i(R_{i,h})} \int_{S_\epsilon}
\frac{dR_i}{R_i-R_{i,h}}.
\eeq
This integral is half the integral over a complete circle taken in a clockwise manner,
\beq
\int\limits_{S_\epsilon} dR_i \frac{\sqrt{1-a_i^2\cF_i}}{\cF_i}  = \frac 1{2\cF'_i(R_{i,h})} \oint_{C_\epsilon}
\frac{dR_i }{R_i-R_{i,h}} = -\frac{i\pi}{2g_{i,h}},
\eeq
where $2g_{i,h} = \cF'_i(R_{i,h})$ is the surface gravity of the horizon. Therefore we find
\beq
{\int\limits_{L_\epsilon}}^{R_{i,h}+\epsilon}dR_i \frac{\sqrt{1-a_i^2\cF_i}}{\cF_i} = 
{\int\limits_{L_\epsilon}}^{R_{i,h}-\epsilon}dR_i \frac{\sqrt{1-a_i^2\cF_i}}{\cF_i} - \frac{i\pi}{2g_{i,h}}.
\eeq
The expression defining the proper time in \eqref{pt} involves a similar integral, but taken over a spatial slice.
Even so, the same argument can be made for this integral (see \cite{aas06,apgs09} for an analogous argument in the 
semi-classical context). If we assume, moreover, that $a(F(r))$ and $P_\Gamma(r)$ 
are both regular across the horizon, the net result is that the phases in the exterior get matched 
to the phases in the interior by the addition of a constant imaginary term, 
\beq
\cW_\text{out}^{(\pm)}(\tau_i,R_i,F_i) = \cW_\text{in}^{(\pm)}(\tau_i,R_i,F_i) \mp \frac{i\pi\omega_i}
{\hbar g_{i,h}}.
\eeq
Thus we can give two independent solutions with support everywhere in the spacetime: an ingoing wave in the 
exterior that is matched to an outgoing wave in the interior
\beq
\psi_i^{(1)}(\tau_i,R_i,F_i) = \left\{\begin{matrix}
e^{\omega_ib_i} \times \exp\left\{-\frac{i\omega_i}\hbar \left[a_i \tau_i + \int^{R_i} dR_i 
\frac{\sqrt{1-a_i^2\cF_i}}{\cF_i}\right]\right\} & \cF_i>0\cr\cr
e^{-\frac{\pi\omega_i}{\hbar{g_{i,h}}}}\times e^{\omega_ib_i} \times \exp\left\{-\frac{i\omega_i}\hbar 
\left[a_i \tau_i + \int^{R_i} dR_i \frac{\sqrt{1-a_i^2\cF_i}}{\cF_i}\right]\right\} & \cF_i < 0 
\end{matrix}\right.
\eeq
and an outgoing wave in the exterior that is matched to an ingoing wave in the interior according to
\beq
\psi_i^{(2)}(\tau_i,R_i,F_i) = \left\{\begin{matrix}
e^{-\frac{\pi\omega_i}{\hbar{g_{i,h}}}}\times e^{\omega_ib_i} \times \exp\left\{-\frac{i\omega_i}\hbar 
\left[a_i \tau_i - \int^{R_i} dR_i \frac{\sqrt{1-a_i^2\cF_i}}{\cF_i}\right]\right\} & \cF_i > 0 \cr\cr
e^{\omega_ib_i} \times \exp\left\{-\frac{i\omega_i}\hbar \left[a_i \tau_i - \int^{R_i} dR_i 
\frac{\sqrt{1-a_i^2\cF_i}}{\cF_i}\right]\right\} & \cF_i < 0
\end{matrix}\right.
\eeq
The first of these solutions represents a flow towards the apparent horizon both in the exterior as well as 
in the interior whereas the second represents a flow away from the apparent horizon in both regions. 
While each solution is self-consistent, neither wave function accurately reflects the physical situation one 
expects from semi-classical collapse, in which an ingoing shell proceeds all the way to the center and the 
horizon emits thermal radiation into the exterior at the Hawking temperature.

To recover this picture we consider a linear superposition of the two wave functions 
\beq
\psi_i = \psi^{(1)}_i + A_i \psi^{(2)}_i,
\eeq
where $A_i$ are complex valued constants. Following \cite{vaz10}, we fix these constants by requiring that 
the current density is constant across the horizon. This implies that $|A_i|^2=1$ and we take $A_i=1$ for 
every shell, which gives an absorption probability of unity for a shell to cross the apparent horizon from 
the exterior. We therefore have
\beq
\psi_i = \left\{\begin{matrix}
\hskip -1.0in e^{\omega_ib_i} \times \exp\left\{-\frac{i\omega_i}\hbar \left[a_i \tau_i + \int^{R_i} 
dR_i \frac{\sqrt{1-a_i^2\cF_i}}{\cF_i}\right]\right\} + & \cr
\hskip 1.0in +~ e^{-\frac{\pi\omega_i}{\hbar{g_{i,h}}}}\times e^{\omega_ib_i} \times \exp\left\{-\frac{i\omega_i}\hbar 
\left[a_i \tau_i - \int^{R_i} dR_i \frac{\sqrt{1-a_i^2\cF_i}}{\cF_i}\right]\right\} & \cF_i > 0 \cr\cr
\hskip -1.0in e^{-\frac{\pi\omega_i}{\hbar{g_{i,h}}}}\times e^{\omega_ib_i} \times \exp\left\{-\frac{i\omega_i}\hbar 
\left[a_i \tau_i + \int^{R_i} dR_i \frac{\sqrt{1-a_i^2\cF_i}}{\cF_i}\right]\right\} + & \cr
\hskip 1.0in +~ e^{\omega_ib_i} \times \exp\left\{-\frac{i\omega_i}\hbar \left[a_i \tau_i - \int^{R_i} dR_i 
\frac{\sqrt{1-a_i^2\cF_i}}{\cF_i}\right]\right\} & \cF_i < 0
\end{matrix}\right.
\label{fullwavefunctions}
\eeq
The second term in the expression for $\psi_i$ in the exterior ($\cF_i>0$) is an outgoing wave that, to an external 
observer, would represent a reflection with relative probability
\beq
\frac{P_{\text{ref},i}}{P_{\text{abs},i}} = e^{-\frac{2\pi\omega_i}{\hbar {g_{i,h}}}}.
\eeq
This is precisely the Boltzmann factor for the shell at temperature $T_{i,H}=\frac{\hbar {g_{i,h}}}
{2\pi k_B}$, where ${g_{i,h}}$ is the surface gravity of the apparent horizon. 

This reflected piece is a purely quantum effect, necessitated by the 
existence of an ingoing wave in the interior, {\it i.e.,} by requiring the continued collapse of the shell
beyond its apparent horizon. This continued collapse is represented by the second term in the expression 
for $\psi_i$ in the interior ($\cF_i<0$). On the other hand, the ingoing wave in the exterior, represented 
by the first term in the expression for $\psi_i$ when $\cF_i>0$, is necessarily accompanied by an outgoing wave 
in the interior occurring with a relative amplitude of $e^{-\frac{\pi\omega_i}{\hbar{g_h}}}$, which is 
equal to the amplitude for ``reflection'' at the apparent horizon. This leads to the alternate picture 
mentioned in the Introduction, in which the Hawking process can be viewed as an effective emission from 
the apparent horizon.

\subsection{Phase Transition}

We can use our results above to examine what happens near the Hawking-Page Transition point \cite{hp83}.
It is worth noting that the Hawking temperature is independent of the energy function. This was first noted
in \cite{vgks07} in connection with non-marginal LTB models without a cosmological constant. Here we have shown 
that the result is robust, holding even in the presence of a negative cosmological constant. 

Now the apparent horizon is given for each shell as the solution of the equation
\beq
2x_{i,h}^{n+1} + n(n+1) x_{i,h}^{n-1} - n(n+1)F_i\Lambda^{\frac{n-1}2} = 0,
\label{hor}
\eeq
where $x_i = R_i\sqrt{\Lambda}$ is dimensionless, and it is straightforward to show that the surface gravity 
for each shell is
\beq
{g_{i,h}} = \frac{\sqrt{\Lambda}}2\left[\frac{2x_{i,h}}n + \frac{n-1}{x_{i,h}}\right].
\label{surfgrav}
\eeq
For fixed $n$ and $\Lambda$,
\beq
\frac{d{g_{i,h}}}{dF_i} = \frac{\sqrt{\Lambda}}2 \frac{dx_{i,h}}{dF_i} \left[\frac 2n - \frac{n-1}{x_{i,h}^2}
\right],
\eeq
so using the fact that $dx_{i,h}/dF_i>0$, which follows directly from \eqref{hor}, we find that $d{g_{i,h}}/dF>0$ 
when $2x_{i,h}^2> n(n-1)$. This is the condition for positive specific heat. When $2x_{i,h}^2 < n(n-1)$ the 
specific heat is negative and the two regimes are separated by the Hawking-Page (phase) transition \cite{hp83}, which 
occurs at $2x_{i,h}^2=n(n+1)$. The wave functions of collapse in \eqref{fullwavefunctions} are
well behaved at the transition point.

\subsection{Wave Functionals}

We now turn to the question of how the collapsing shell wave functions described in the previous section may be combined 
to yield wave functionals. Obviously the superposed wave functions of \eqref{fullwavefunctions} cannot be directly 
used for this purpose as they are not the simple exponentials required for diffeomorphism invariance by 
\eqref{wavefunctionaldef}. Instead, we take the continuum limit of the product of the shell wave functions 
$\psi_i^{(1)}$ and $\psi_i^{(2)}$ separately to form two corresponding diffeomorphsim invariant wave functionals, 
$\Psi_1$ and $\Psi_2$. Then, we take a linear combination of these wave functionals to form the full wave functional describing the collapse \cite{vaz10}	.

Accordingly, the functional equivalent of the superposed wave functions in \eqref{fullwavefunctions} is
$\Psi=\Psi_1+\Psi_2$, or
\beq
\Psi= \left\{\begin{matrix}
\hskip -0.5in e^{\frac 12\int dr \Gamma b} \times \exp\left\{-\frac i{2\hbar} \int^R dr \Gamma \left[a\tau + 
\int_{r=\text{const.}}^{R} dR \frac{\sqrt{1-a^2\cF}}{\cF}\right]\right\} + & \cr\cr
\hskip 0.5in +~ e^{-S/2}\times e^{\frac 12\int dr \Gamma b} \times \exp\left\{-\frac i{2\hbar} 
\int dr \Gamma \left[a \tau - \int_{r=\text{const.}}^R dR \frac{\sqrt{1-a^2\cF}}{\cF}\right]\right\} & \cF > 0 \cr\cr\cr
\hskip -0.5in e^{-S/2}\times e^{\frac 12\int dr\Gamma b} \times \exp\left\{-\frac 
i{2\hbar} \int dr \Gamma \left[a \tau + \int_{r=\text{const.}}^R dR \frac{\sqrt{1-a^2\cF}}{\cF}\right]\right\} + & \cr\cr
\hskip 0.5in +~ e^{\frac 12\int dr \Gamma b} \times \exp\left\{-\frac i{2\hbar}\int dr \Gamma \left[a \tau - 
\int_{r=\text{const.}}^R dR \frac{\sqrt{1-a^2\cF}}{\cF}\right]\right\} & \cF < 0
\end{matrix}\right.
\eeq
When use is made of \eqref{hor} and \eqref{surfgrav}, the relative amplitude, $e^{-\frac\pi{2\hbar} 
\int dr \Gamma/{g_h}}$, works out to precisely $e^{-S/2}$, where $S=A_h/4\hbar G_d$ is the 
Bekenstein-Hawking entropy of the black hole. Thus the ratio of the reflection probability to the 
probability for absorption is determined only by the entropy of the black hole,
\beq
\frac{P_\text{ref}}{P_\text{abs}} = e^{-S}
\eeq
and we have recovered the results of \cite{vaz10} in the more general setting of $d-$dimensional, non-marginal 
collapse and in the presence of a (negative) cosmological constant.

\section{Conclusions}

In this paper we have used the wave-functionals of an exact midi-superspace quantization of the 
non-marginally-bound LTB models in the presence of a cosmological constant and in an arbitrary number of 
spatial dimensions to study the Hawking evaporation process.  As in previous works, regularization was 
performed on a lattice and the wave functionals were shown to be constructed out of wave functions 
describing individual shells of collapsing dust. The apparent horizon is an essential singularity of 
the Wheeler-DeWitt equation and the solutions of the latter could only be given by quadrature separately 
in the exterior and in the interior of the apparent horizon. The central issue discussed here was how 
the interior and exterior solutions can be matched across the horizon. To accomplish the required matching 
we defined the integrals appearing in the solution of the Wheeler-DeWitt equation 
by deforming the integration path in the complex plane to go around the pole at the apparent horizon. This 
implied that crossing the horizon involved a rotation of the dust proper time in the complex plane and had 
the effect of introducing an imaginary constant into the phase of the outgoing wave 
functions. We were then able to show that an ingoing shell wave function in one region is required to be accompanied by an outgoing shell wave function in the other region. The relative amplitude of the outgoing 
wave function in each case was shown to be given by the square root of the Boltzamann factor at the Hawking
temperature appropriate to the shell.  

The approach in this paper enjoys several advantages over the approach via Bogoliubov coefficents, while also 
producing an alternative and attractive view of the evaporation process. In the first place, no near horizon 
expansion of the wave-functional is necessary. Secondly, the approach via Bogoliubov coefficients does not 
get much beyond the semi-classical level because it is necessary to approximate the mass function in 
such a way that it represents a massive black hole surrounded by tenuous dust. Hence one is effectively 
looking at the semi-classical radiation from the event horizon of a static black hole. By contrast, no such 
approximation to the mass function is necessary here, so we are genuinely examining the radiation from the 
apparent horizon {\it during} collapse. Thirdly, the inner product used in the calculation of the Bogoliubov 
coefficient is not the one that is uniquely determined by the lattice regularization (see Appendix B of 
\cite{kmv06}) but one that is determined from the DeWitt supermetric. This can be justified only in 
the approximation described above because, for this case alone, no measure is uniquely determined by the 
regularization scheme \cite{vw09}). In all other cases, the measure is uniquely determined and different 
from that provided by the DeWitt supermetric. The results we report here are independent of the inner product.

If the matter is assumed to undergo continued collapse we showed that the relative probability for the shell 
wave function to cross the apparent horizon is unity, whether it is incident from the interior or the exterior.
This was argued to lead to two pictures of the evaporation process. In the first picture, one takes the point of 
view of an external observer with no access to the interior. To this observer the horizon appears to possess a non-zero
reflectivity. On the other hand, the observer who has access to the entire wave function sees the outgoing 
exterior portion of the shell wave function as a transmission of an outgoing interior wave across the horizon,
which exists because of the collapsing exterior. Thus the horizon can also be thought of as an emitter.

We showed that the Hawking temperature is independent of the energy function, {\it i.e.,} of the initial velocity 
distribution of the shells, and that the shell wave functions are well behaved at the Hawking-Page transition
point during the collapse. Moreover, the relative amplitude for outgoing wave functionals is $e^{-S/2}$, where 
$S$ is the Bekenstein-Hawking entropy of the final state black hole even in the presence of the cosmological 
constant. This generalizes \cite{vaz10}, for which closed form solutions were available. The results are therefore 
generic to dust collapse.
\bigskip\bigskip

\noindent{\bf Acknowledgements}
\bigskip

\noindent C.V. is grateful to T.P. Singh and to the Tata Institute of Fundamental Research 
for their hospitality during the time this work was completed. K.L. and C.V. acknowledge useful conversations with
T.P. Singh and the partial support of the Templeton Foundation under Project ID $\#$ 20768.

\end{document}